\documentclass{aa}

\usepackage{amsmath,amssymb}
\usepackage{graphicx}

\begin{document}

\thesaurus{02.07.1; 02.08.1; 02.09.1; 03.13.1; 12.03.4; 12.12.1}

\title{Extending the Scope of Models for Large--Scale Structure 
Formation in the Universe}
\author{Thomas Buchert\inst{1,2}, Alvaro Dom\'\i nguez\inst{3}, and Juan
  P\'erez-Mercader\inst{3}}
\offprints{A. Dom\'\i nguez}
\institute{Theory Division, CERN, CH-1211 Geneva 23, Switzerland
\and
  On leave from: Theoretische Physik, Ludwig--Maximilians--Universit\"at,
  Theresienstr. 37, D--80333 M\"unchen, Germany
  (e-mail:buchert@theorie.physik.uni-muenchen.de)
\and
Laboratorio de Astrof\'\i sica Espacial y F\'\i sica Fundamental,
Apartado 50727, E--28080 Madrid, Spain \\
(e-mails:dominguez@laeff.esa.es, mercader@laeff.esa.es)}

\date{}

\authorrunning{Buchert, Dom\'\i nguez, P\'erez-Mercader}
\titlerunning{Extending the Scope of Models for LSS formation}

\maketitle

\begin{abstract}

We propose a phenomenological generalization of the models of
large--scale structure formation in the Universe by gravitational
instability in two ways: we include pressure forces to model
multi--streaming, and noise to model fluctuations due to neglected
short--scale physical processes. We show that pressure gives rise to a
viscous--like force of the same character as that one introduced in
the ``adhesion model'', while noise leads to a roughening of the
density field, yielding a scaling behavior of its correlations.


\end{abstract}

\section{Introduction}

Analytical approximations for the evolution of large--scale structure
(LSS) are based on the paradigm that small initial perturbations grow
by gravitational instability, which is in turn implemented in the
simplest matter model, `dust' (Peebles 1980, Zel'dovich \& Novikov
1983, Sahni \& Coles 1995, and ref.  therein). However, this
approximation has some limitations: one has to restrict the
application to the early stages of structure formation and when the
effects on the evolution of physical processes different from
gravitational instability are negligible. In this paper we purport to
generalize this matter model in order to overcome some of these
limitations.

\medskip

One of the problems at the later stages of LSS evolution is the
formation of multi--stream regions, i.e., regions where particles of
dust come together with very different velocities. This fact manifests
itself as the emergence of caustics in the density field, where the
velocity field is ``vertical'' (i.e., where it acquires an infinite
derivative at a point) and later on multiply valued. This problem
arises from insisting on following the trajectory of each particle of
dust. We therefore propose a set of hydrodynamic--like equations for
the {\em coarse--grained} fields, which trace the average motion
rather than that of individual particles. A substantial ingredient of
this approach is that the coarse--grained velocity evolves under the
combined action of gravity and {\em pressure--like} forces due to
velocity dispersion (i.e., because particles of dust do not move
exactly with the coarse--grained velocity). We cannot resort to
hydrodynamic considerations of local equilibrium but make instead use
of `equations of state' as phenomenological matter models without any
further justification. In these models the pressure is assumed
isotropic and may only depend on the (coarse--grained) density, that
is, $p=p(\varrho)$.  This assumption makes the problem accessible to
analytical study and helps to illuminate our major argument that the
presence of forces which counteract the gravitational attraction is a
basic step in understanding the dynamics of self--gravitating matter;
a detailed study of the origin and properties of pressure forces is
carried out elsewhere (Buchert \& Dom\'\i nguez 1998).

\medskip

That the putative problem is far more complex than the dust case
already becomes clear in the investigation of the one--dimensional
Euler--Newton system with the simple matter model $p \propto \varrho$
(G\"otz 1988). G\"otz has shown that solutions to the one--dimensional
problem can be generated by solutions of the Sine--Gordon equation.
This well--studied equation has a rich spectrum of solutions that
includes solitons. G\"otz also pointed out that an asymptotic
N--soliton state is generic, i.e., will be realized almost
independently of the initial data. We see already in this
comparatively simple case, that we are faced with a generic picture
which is completely different from what emerges in a cosmology based
on dust matter: special nonlinear features build up structures at
large times which are {\it absent} in the dust cosmology. This
illustrates that the complexity introduced by a pressure term could
bear far--reaching surprises. We also want to stress that the
introduction of a pressure term is not in contradiction with our
present understanding of LSS formation, since it partly arises from
N--body simulations of the structure formation process which capture
multi--streaming effects, and hence are not constrained by the
analytical approximation to dust matter\footnote{We here imply
  `single--streamed dust' as opposed to models which also cover
  multi--dust regions.}  which, in its simplest realizations (e.g.,
Zel'dovich's approximation (Zel'dovich 1970)), features immediate
decay of structures after their formation.

\medskip

We also consider an extension of the model to include {\em stochastic}
effects. This ``stochasticity'' arises from the effect on the
dynamical evolution of physical processes occurring on time-- and/or
length--scales much smaller than those directly associated with LSS
formation, thus allowing to model them by means of a stochastic source
(a noise). Possible sources are deviations from the mean field
approximation, fluctuations inherent to the hydrodynamic (i.e.,
coarse--grained) description, and non--gravitational processes in
baryonic matter. We shall use the simplest model of a
Gaussian--distributed forcing on the coarse-grained evolution.

\medskip

As with pressure--like forces, we just want to stress that a noisy
forcing could be relevant to LSS formation, but a detailed
consideration of its origin and properties is beyond the scope of the
present paper. In fact, application of the Renormalization Group shows
that noise is relevant in non--exceptional conditions (Barbero et al.
1997; Dom\'\i nguez et al. 1999), implying that even if it is very
weak (apparently negligible), its effects are amplified and can have a
non--negligible effect on the dynamical evolution.

\medskip

This paper is structured as follows: in Sect. \ref{basic_equations} we
begin by presenting the basic system of equations in the Newtonian
regime, and then proceed to a discussion of restrictive assumptions
for the weakly nonlinear regime. In Sect. \ref{pressure} we discuss
the role of the pressure--like force for some particular choices of
the equation of state $p=p(\varrho)$ and the connection with Burgers'
equation. In Sect. \ref{KPZ} we consider the role of noise and provide
a detailed description of the relationship between the cosmological
equations and the Kardar--Parisi--Zhang (KPZ) equation. In Sect.
\ref{linear_regime} we study the linear regime in the presence of
pressure and noise. We finally conclude in Sect. \ref{conclusions}.
Some technicalities have been left for two appendices, one devoted to
the exploration of the validity of what we call in Sect. \ref{KPZ} the
``adiabatic approximation'', and another to a more detailed discussion
of the linear regime.

\bigskip

\section{Basic Equations and Restrictive Assumptions} \label{basic_equations}

We are interested in discussing LSS formation in the non--relativistic
regime and therefore consider the Newtonian cosmological equations for
a self--gravitating fluid in a standard Friedmann--Lema\^\i tre (FL)
cosmological background dominated by non--relativistic matter (Peebles
1980). The cosmological background is characterized by the cosmic
expansion factor $a(t)$ and the homogeneous background matter density
$\varrho_{b}(t)$, which obey (Hubble's function is defined as $H =
\dot{a} / a$)
\begin{equation}
H^2 = \frac{8 \pi G}{3} \varrho_{b} - \frac{K}{a^2} + 
\frac{\Lambda}{3} \; , \;\;\;\;\; \varrho_{b} = \varrho_{0} a^{-3} \;\; ,
\end{equation}

\noindent
where the constant $K$ determines the sign of the spatial curvature
(treated as an integration constant in the Newtonian framework
considered throughout this paper), $\varrho_{0}$ is the background
density at some time $t_{0}$ when $a(t_{0})=1$, and $\Lambda$ is the
cosmological constant. Without loss of generality, one can choose
$t_{0}$ to correspond to the present epoch.

It is convenient to work in comoving coordinates ${\bf x} \equiv
a^{-1}{\bf r}$, where $\bf {r}$ are the standard non--rotating
Eulerian coordinates. The fundamental fields will be as follows: the
density $\varrho$ (or equivalently, the density contrast $\delta :=
(\varrho / \varrho_{b}) - 1$ ), the peculiar--velocity ${\bf u} \equiv
{\bf u}_{phys} - H {\bf r}$, where ${\bf u}_{phys}$ is the physical
velocity and $H {\bf r}$ is the Hubble flow, and the gravitational
peculiar--acceleration ${\bf g} = {\bf g}_{phys} + \frac{1}{3}\left(4
  \pi G \rho_{b} - \Lambda \right) {\bf r}$, where ${\bf g}_{phys}$ is
the physical gravitational acceleration and $-\frac{1}{3}\left( 4 \pi
  G \varrho_{b} - \Lambda \right) {\bf r}$ is the Newtonian
counterpart of the gravitational acceleration opposing to the
background expansion. We subject $\varrho$, ${\bf u}$ and ${\bf g}$ to
periodic boundary conditions on some large scale to assure uniqueness
of the cosmological solutions, in which case $\varrho_{b}$ is equal to
the spatially averaged density (see Ehlers \& Buchert 1997).

The fields $\varrho$, ${\bf u}$ obey a set of hydrodynamic equations
expressing the conservation of mass and momentum in an expanding
background. These equations are quite similar to those of the standard
dust model (Peebles 1980), except for two forcing terms in Euler's
equation that model the multi-streaming and stochastic effects
discussed in the Introduction \footnote{From now on, time derivatives
are taken at constant ${\bf x}$ and gradients refer to comoving
coordinates.}: \newline

\noindent
$\bullet$ Continuity equation:
\begin{equation}
\frac{\partial \varrho}{\partial t} + 3 H \varrho +
\frac{1}{a} \nabla \cdot (\varrho {\bf u}) = 0 \; ; 
\label{cosmohydroa}
\end{equation}
$\bullet$ Euler's equation:
\begin{equation}
\frac{\partial {\bf u}}{\partial t} + \frac{1}{a} 
({\bf u} \cdot \nabla){\bf u} + H {\bf u} = {\bf g} - 
\frac{1}{a \varrho} \nabla p + {\bf s} \; ;
\label{cosmohydrob}
\end{equation}
$\bullet$ Newtonian field equations:
\begin{equation}
\nabla \cdot {\bf g} = - 4 \pi G a (\varrho - \varrho_{b}) 
\; , \;\;\;\;\; \nabla \times {\bf g} = 0 \; .
\label{cosmohydroc}
\end{equation}
 
We emphasize that the integral curves of the peculiar--velocity field
${\bf u}$ are not associated to trajectories of {\it individual}
particles, rather $\varrho$, ${\bf u}$ and ${\bf g}$ are considered as
coarse--grained fields. This coarse--graining is the origin of the two
new terms on the right--hand--side of Euler's equation
(\ref{cosmohydrob}). One of these new terms is the pressure force
$\nabla p$, which accounts for the isotropic part of the multi--stream
force, and therefore models velocity dispersion (that is, the fact
that in any infinitesimal cell there are particles with different
velocities). Because of this, the integral curves of ${\bf u}$
represent trajectories of the {\it mean} (possibly multi--streamed)
flow after averaging over velocity space. This pressure term is {\em
  not} related to thermal pressure, which can be indeed neglected on
the scales we are interested in. It is a model of the velocity
dispersion generated by gravitational instability (see Buchert \&
Dom\'\i nguez 1998 and, for a recent generalization to general
relativity, Maartens et al. 1999).
 
The other new term is the stochastic force represented by the noise
${\bf s}$; it accounts for processes hidden by the coarse--grained
description of the fluid and whose typical time-- and length--scales
are much shorter than those explicitly considered for LSS formation.
We have resorted to modelling these processes as a stochastic forcing
and include: (a) the effects of small--scale degrees of freedom whose
physics is also governed by non--gravitational processes, (b)
deviations from mean field behavior, manifested as random forces
acting on the particles of the gravitational gas as a consequence of
independent impulses of random size and amplitude arising from
``sling-like'' processes in encounters (Kandrup 1980), (c) deviations
of the density and velocity fields from the values prescribed by the
deterministic version of Eqs. (\ref{cosmohydroa}-\ref{cosmohydroc})
due to the graininess of the underlying physical system of particles
(Lifshitz \& Pitaevskii 1980).

To close the system of Eqs. (\ref{cosmohydroa}-\ref{cosmohydroc}) a
relation is needed between the dynamical pressure $p$ and the two
independent fields $\varrho$ and ${\bf u}$, as well as a specification
of the statistical properties of the stochastic force ${\bf s}$. As
for the former, we assume the local relationship $p=p(\varrho)$.
There does not seem to be any {\sl a priori} reason for this
``slaving'' of pressure to density (one cannot resort to the
hypothesis of local equilibrium as is done for the thermodynamical
pressure in fluids), so the success of this {\em phenomenological}
assumption must be judged according to the conclusions following from
it. In fact, a detailed study of the origin of pressure forces in Eq.
(\ref{cosmohydrob}) provides $p \propto \varrho^{5/3}$ under the
assumption of small velocity dispersion (Buchert \& Dom\'\i nguez
1998) and therefore $p=p(\varrho)$ is the most straightforward
phenomenological generalization. We also require $p'({\varrho})>0$,
that is, pressure opposes gravitational collapse.

As for the noise, we make the assumption of Gaussian distributed
noise. Since noise is due to the short--scale degrees of freedom, this
assumption could be justified by the central limit theorem. As is well
known, Gaussian noise can be characterized by just two moments: its
mean, which we require to vanish, $\langle {\bf s} \rangle = {\bf 0}$
(since any systematic forcing should be made explicit in Eq.
(\ref{cosmohydrob})), and its two--point correlations
\begin{equation}
\langle s_{i}({\bf x}, t) s_{j}({\bf x}', t') \rangle = 2 D_{ij} 
({\bf x}, {\bf x}', t, t') \; ,
\label{scorrelation}
\end{equation}

\noindent
where $D_{ij} ({\bf x}, {\bf x}', t, t')$ is the covariance matrix
with mixed discrete and continuous indices (Gardiner 1994; Van Kampen
1992). We ignore the possibility that $D_{ij}$ depends on $\varrho$ or
${\bf u}$ (quenched noise), since this renders the analysis of Eqs.
(\ref{cosmohydroa}-\ref{cosmohydroc}) too difficult. As with pressure,
this must be seen as a {\em phenomenological} assumption, whose merit
will be judged from the final results. Notice however that we consider
the possibility of {\em colored} Gaussian noise, i.e., the noise $s$
is correlated over space and time, as expressed by the (in general)
non--trivial dependence of $D$ on position and time in Eq.
(\ref{scorrelation}). Colored noise represents a richer texture of
physical effects than white noise and the particular case of
power--law correlated noise is still amenable to analytical study by
means of the Renormalization Group (Dom\'\i nguez et al. 1999). We
later restrict the generality further by choosing a curl--free noise
and characterize it statistically by a single function $D$ instead of
$D_{ij}$.

\medskip The analytical study of the system of Eqs.
(\ref{cosmohydroa}-\ref{cosmohydroc}) is very difficult in general.
Inspired by the deterministic dust case (i.e., $p=0$, ${\bf s}={\bf
  0}$), we simplify the problem and put forward the assumption of
parallelism: we impose the condition that the peculiar--velocity is a
potential field and remains parallel to the gravitational
peculiar--acceleration field:
\begin{equation}
{\bf g}= F(t) {\bf u} \; ,
\label{parallelism}
\end{equation}

\noindent
where $F(t)>0$ is a proportionality coefficient which follows from the
deterministic linear theory for dust (see Sect. \ref{linear_regime}
and in particular the discussion after Eqs.
(\ref{Fexpansiona}-\ref{Fexpansionb}) in Appendix B). This assumption
implicitly requires that both pressure and the intensity of the noise
are small when compared with the dominant self--gravity in Eq.
(\ref{cosmohydrob}), which restricts our considerations to spatial
scale regimes close to the validity limit of the dust model; noise and
pressure will typically dominate on scales small compared to this
limit. The assumption of parallelism underlies the well--known
``Zel'dovich approximation'' (Zel'dovich 1970), and is well--justified
for deterministic dust models in the linear as well as in the weakly
nonlinear regimes (see Bildhauer \& Buchert 1991; Kofman 1991; Buchert
1992; Hui \& Bertschinger 1996; Susperregi \& Buchert 1997). This
(indeed oversimplifying assumption) will be very useful to
analytically access the problem, and to define ``local''
approximations.

Also, we study that assumption first because popular models like the
``adhesion approximation'' (Gurbatov et al. 1989) can be {\it derived}
on the basis of this assumption. As we shall see, this assumption is
consistent with the picture that, coming from large scales, the mean
motion is ruled by the dust model, but incorporation of the effects of
pressure and noise adds several interesting aspects to it. At the
smaller scales, the ``parallelism assumption'' requires that the
``backreaction'' of these effects on the trajectories of the mean dust
flow be neglected. In the dust case it can be shown that this
assumption admits a class of 3D solutions (Buchert 1989); this class
is highly restrictive, but using it as the basis for approximation
schemes turns out to be very successful (e.g. Buchert 1996 and ref.
therein) - a subcase of these schemes is known as Zel'dovich's
approximation. However, in the present case we cannot disprove that
the assumption (\ref{parallelism}) is too restrictive to allow for the
existence of a class of exact solutions. We will learn more about the
justification of this assumption in Sect. \ref{linear_regime}.

\medskip

Under the assumption (\ref{parallelism}), Eq. (\ref{cosmohydrob})
reduces to
\begin{equation}
\frac{\partial {\bf u}}{\partial t} + \frac{1}{a} ({\bf u} \cdot \nabla)
{\bf u} + (H-F) {\bf u} = - \frac{p'(\varrho)}{a \varrho} \nabla \varrho 
+ {\bf s}  \; ,
\label{euler2}
\end{equation}

\noindent
with constraints following from the field equations
(\ref{cosmohydroc}): the velocity field is irrotational, $\nabla
\times {\bf u} = {\bf 0}$, and its divergence satisfies
\begin{equation}
F \nabla \cdot {\bf u} + 4\pi G a (\varrho - \varrho_b )= 0 \; .
\label{constraint}
\end{equation}

\noindent
The pressure force term $\nabla \varrho$ can be computed from this
last expression, which gives
\begin{equation}
\nabla \varrho \; = - \frac{F}{4 \pi G a} \nabla ^2 {\bf u} \;,
\end{equation}

\noindent
and Euler's equation may now be finally written as:
\begin{equation}
\frac{\partial {\bf u}}{\partial t} + \frac{1}{a} ({\bf u} \cdot \nabla)
{\bf u} + (H-F) {\bf u} = \nu \nabla^2 {\bf u} + {\bf s} \; ,
\label{eulerapp}
\end{equation}

\noindent
where we have defined a coefficient $\nu$ which behaves as a kinematic
viscosity and which we call {\em ``gravitational multi--stream''} (GM)
coefficient (because it has its origin in the interplay between
self--gravitation and multi--streaming flow); it is given by
\begin{equation}
\nu : = \frac{F(t) p'(\varrho)}{4 \pi G a(t)^2 \varrho} \; > \; 0 \;, 
\label{GMviscosity}
\end{equation}

\noindent
and depends on the density, and explicitly on time through $F(t)$ and
$a(t)$. As we see from Eqs. (\ref{eulerapp}) and (\ref{GMviscosity}),
the pressure behaves in Euler's equation {\em effectively} as a
viscous force that prevents caustics in the velocity field, a behavior
that essentially requires the participation of self--gravity. The
interplay between self--gravity and pressure leads to the
stabilization of the structures, and holds them together. We also want
to stress that the GM coefficient {\em does not} generate dissipation,
since our starting set of Eqs. (\ref{cosmohydroa}-\ref{cosmohydroc})
lacks any sort of dissipative term (see Buchert \& Dom\'\i nguez 1998
for further details about this apparently paradoxical point).

The last step is the elimination of the density $\varrho$ in the
expression of the GM coefficient in favor of $\nabla \cdot {\bf u}$ by
means of the constraint (\ref{constraint}), thus reducing Eq.
(\ref{eulerapp}) to a {\em closed} equation for the velocity field
${\bf u}$. This equation will be explored in the next sections.

\bigskip

\section{The Role of Pressure: Burgers--like Equations} \label{pressure}

\begin{figure}
  \resizebox{\hsize}{!}{\includegraphics[width=9cm]{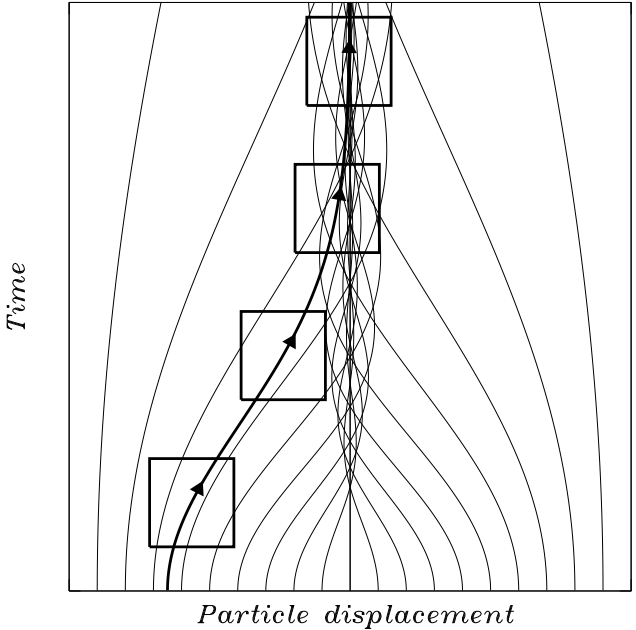}}
  \caption{The coarse-graining idea is exemplified in a schematic way based
    on a 2D tree-code simulation (Buchert 1996): the particles are
    attracted towards the pancake, a multi-stream region containing
    many streams (velocities) at a given Eulerian position. The
    coarsening cell follows initially a trajectory similar to that of
    individual particles. But as the cell moves into the pancake, the
    kinetic energy of the bulk motion is gradually transformed into
    internal kinetic energy, and its trajectory becomes qualitatively
    different from that of a particle.}
  \label{multistream}
\end{figure}

In this section we consider the role of pressure in the dynamical
evolution and thus for the time being we drop the noise term, ${\bf
  s}({\bf x}, t)$. But before going into the details, it is worthwhile
to get an intuitive picture of how the pressure term affects the
dynamical evolution. As remarked above, pressure has in Eq.
(\ref{eulerapp}) a stabilizing effect opposing collapse. But this
effect holds even in regimes where the assumptions leading to Eq.
(\ref{eulerapp}) (especially the parallelism assumption) break down:
Fig. \ref{multistream} shows the evolution of coarsening cells in a
simulation and how velocity dispersion influences their motion.

Different relationships between pressure and density yield different
$\varrho$--dependences of the GM coefficient. In this section we will
concentrate on polytropic models $p = \kappa \varrho^{\gamma}$, where
$\kappa$ is constant and $\gamma$ is the polytropic index. The
prefactor $\kappa$ should be small enough so that the parallelism
assumption (\ref{parallelism}) is approximately satisfied on the
length scales we are interested in, and it should also satisfy $\kappa
\gamma >0$ in order to fulfill the condition $p'(\varrho)>0$. The
choice of polytropic models is interesting because we are able to
recover well--known approximations in cosmology (Zel'dovich's
approximation, the ``sticky particle model'' and the ``adhesion
model'') as well as other interesting models, and we can understand
them within a single equation. With this form of $p(\varrho)$, Eq.
(\ref{eulerapp}) becomes
\begin{equation}
\frac{\partial {\bf u}}{\partial t} + \frac{1}{a} ({\bf u} \cdot \nabla)
{\bf u} + (H-F) {\bf u} = \frac{\gamma \kappa F}{4 \pi G a^2}
\varrho^{\gamma-2} \nabla^2 {\bf u} \; .
\label{polytropic}
\end{equation}

\noindent
We employ now Eq. (\ref{constraint}) to eliminate $\varrho$ in place
of $\nabla \cdot {\bf u}$, and since $\nabla^2 {\bf u} = \nabla
(\nabla \cdot {\bf u})$ (because $\nabla \times {\bf u} = {\bf 0}$)
one gets \footnote{Eq. (\ref{closed_for_u}) is valid for $\gamma \neq
  1$.  If $\gamma = 1$, one has a logarithmic function instead of a
  power on the right--hand--side of Eq. (\ref{closed_for_u}), see Eq.
  (\ref{isothermal}).}:
\begin{equation}
\frac{\partial {\bf u}}{\partial t} + \frac{1}{a} ({\bf u} \cdot \nabla)
{\bf u} + (H-F) {\bf u} = -{\tilde \nu} \nabla [1 - \beta \nabla \cdot 
{\bf u}]^{\gamma - 1} \; ,
\label{closed_for_u}
\end{equation}

\noindent
where ${\tilde \nu} = \gamma \kappa \varrho_{b}^{\gamma - 1} /
(\gamma-1) a$ and $\beta = F / 4 \pi G a \varrho_{b}$ are functions
which depend only on time. This equation can be further simplified as
follows: making explicit use of $b(t)$, the growing mode of the
density field in the linear regime of the deterministic dust case, one
has $F=4 \pi G \varrho_{b} b/{\dot b}$ (see Eqs.
(\ref{Fexpansiona}-\ref{Fexpansionb}) in Appendix B and discussion
thereafter). Hence, defining a new velocity field ${\bf v} = {\bf u} /
a {\dot b}$ and employing $b$ as the new time variable, Eq.
(\ref{closed_for_u}) can be written as:
\begin{equation}
\frac{\partial {\bf v}}{\partial b} + ({\bf v} \cdot \nabla)
{\bf v} = -\mu \nabla [1 - b \nabla \cdot {\bf v}]^{\gamma - 1} 
\; , \;\;\;\;\; \mu := \frac{\tilde {\nu}}{a {\dot b}^2} \; .
\label{burgers}
\end{equation}

\noindent
When $\gamma = 2$, this equation becomes the 3D Burgers' equation
(Burgers 1974) (except for the time--dependence of the GM
coefficient $\mu$). In the cases $\gamma \neq 2$, we are dealing with
generalizations of Burgers' equation. We now discuss several cases for
the choice of $\gamma$ \footnote{The names given below to some of the models
stem from thermodynamical notions; we kept those names but do not
imply that we describe situations in thermodynamical equilibrium.}.

\bigskip\medskip
\noindent
\centerline{\it Zel'dovich's approximation}

\medskip
If one chooses $\gamma=0$, then $\mu = 0$ and Eq. (\ref{burgers}) 
reduces to 
\begin{equation}
\frac{\partial {\bf v}}{\partial b} + ({\bf v} \cdot 
\nabla){\bf v} = 0 \; ,
\label{zeldovich}
\end{equation}

\noindent
which corresponds to the well--known Zel'dovich's approximation; that
approximation is equivalent to a subclass of Lagrangian first--order
solutions, the subclass being just defined by our assumption
(\ref{parallelism}) (see Buchert 1989, 1992). The dynamical evolution
governed by this equation generically leads to the formation of
singularities where the velocity field is multi--valued (Arnol'd et
al. 1982).

\newpage
\noindent
\centerline{\it Adhesion approximation}

\medskip
With the choice $\gamma=2$, the GM coefficient is independent of the
density, and Eq. (\ref{burgers}) becomes Burgers' equation with a
time--dependent ``viscosity'':
\begin{equation}  
\frac{\partial {\bf v}}{\partial b} + ({\bf v} \cdot 
\nabla){\bf v} = \mu b \nabla^2 {\bf v} \; .
\label{adhesion}
\end{equation}

\noindent
This equation is formally equivalent to the so--called ``adhesion
approximation'' (Gurbatov et al. 1989; Weinberg \& Gunn 1989), aside
from the time--dependence of the GM coefficient \footnote{In an
  Einstein--de Sitter background cosmology $b \mu \approx b^{-3}$ and
  thus the GM coefficient in (\ref{adhesion}) approaches zero as time
  goes by. Remember also that in the ``adhesion approximation'' the
  constant coefficient was introduced {\it ad hoc} to
  phenomenologically model gravitational ``sticking of particles'' as
  observed in N--body simulations.}.

Letting $\kappa \rightarrow 0$ (which implies $\mu \rightarrow 0$) in
this model, we get ``maximal adhesion'' and the large--scale structure
is built from a skeleton of the ``honeycomb--type'': the evolution is
governed by Zel'dovich's approximation, Eq. (\ref{zeldovich}),
everywhere except at the singularities, which become shock fronts.
This limiting model is known as ``sticky particle model'', for which
geometrical contruction methods have been developed (Pogosyan 1989;
Kofman et al. 1992; for review see Sahni \& Coles 1995). In our
derivation this limit implies $p \rightarrow 0$, i.e., vanishing
velocity dispersion. It must be noticed, however, that this limit is
singular: if $p = 0$ {\em strictly}, then we would recover
Zel'dovich's approximation, which is qualitatively different from the
``sticky particle model''.  Therefore, a non--vanishing, though very
small, velocity dispersion is required to recover typical properties
of the ``adhesion model''.

\bigskip\medskip
\centerline{\it Isothermal model}

\medskip
This model corresponds to $\gamma = 1$ and is mostly studied in
connection with the linear theory of gravitational instability (since
linearization of the pressure term in Eq. (\ref{cosmohydrob}) amounts to
choosing $\gamma=1$). Eq. (\ref{polytropic}) yields:
\begin{equation}  
\frac{\partial {\bf v}}{\partial b} + ({\bf v} \cdot 
\nabla){\bf v} = - \frac{\kappa}{a^2 {\dot b}^2} \nabla 
\ln (1 - b \nabla \cdot {\bf v}) \; .
\label{isothermal}
\end{equation}

\noindent
It is worth mentioning that Eq. (\ref{polytropic}) particularized
to $\gamma = 1$ resembles the Navier--Stokes equation with respect to the
density dependence of the GM coefficient.

As pointed out in the introduction, G\"otz (1988) has shown that the
asymptotic state of this model in one--dimensional space and
neglecting the expanding background is a collection of solitons. On
the other hand, parallelism is not an approximation but an exact
relationship in one dimension (although the proportionality
coefficient will be in general a function of time {\it and} position).
Hence, one can expect that Eq. (\ref{isothermal}) leads to a similar
picture to that obtained from the adhesion model, which is in turn
consistent with that offered by G\"otz: the shock fronts, stabilized
by velocity dispersion, will play the role of solitons.

\bigskip\medskip
\centerline{\it Adiabatic model}

\medskip
For this model $\gamma=5/3$ and Eq. (\ref{burgers}) becomes
\begin{equation}
\frac{\partial {\bf v}}{\partial b} + ({\bf v} \cdot \nabla)
{\bf v} = - \mu \nabla [1 - b \nabla \cdot {\bf v}]^{2/3} \; .
\end{equation}

\noindent
The interesting point about this choice is that there are physical
reasons to prefer this model over any other one in the early stages of
LSS formation: Buchert \& Dom\'\i nguez (1998) show how this form of
$p(\varrho)$ can be {\em derived} from dynamical considerations.

\bigskip\medskip
\centerline{\it Cosmogenetic model}

\medskip
By choosing $\gamma=4/3$ we recover the ``cosmogenetic'' model
(Chandrasekhar 1967):
\begin{equation}
\frac{\partial {\bf v}}{\partial b} + ({\bf v} \cdot \nabla)
{\bf v} = - \mu \nabla [1 - b \nabla \cdot {\bf v}]^{1/3} \; .
\end{equation}

\noindent
This model is of interest because it is the only polytropic model
compatible with ``comoving hydrostatic equilibrium'', i.e., with the
solutions of Eqs. (\ref{cosmohydroa}-\ref{cosmohydroc}) corresponding
to ${\bf u} \equiv {\bf 0}$, so that the temporal evolution of the gas
simply follows the expansion: Eq. (\ref{cosmohydroa}) yields
$\varrho({\bf x}, t) = a^{-3}(t) \varrho_{0}({\bf x})$, Eq.
(\ref{cosmohydroc}) then becomes ${\bf g}({\bf x}, t) = a^{-2}(t) 
{\bf g}_{0} ({\bf x})$. The ``hydrostatic equilibrium'' condition follows
from Eq. (\ref{cosmohydrob}) (dropping the noise): $\nabla p = a
\varrho {\bf g}$. Combining this condition with the polytropic
relationship yields
\begin{equation}
\gamma \kappa a^{-3 \gamma} \varrho_{0}^{\gamma - 1} \nabla \varrho_{0} 
= a^{-4} \varrho_{0} {\bf g}_{0} \;\; ,
\end{equation}
which can be satisfied only if $\gamma=4/3$. (Obviously, the state of
``comoving hydrostatic equilibrium'' itself is incompatible with the
parallelism assumption).

\bigskip
\bigskip

To conclude this section, one can conjecture that Burgers--like
equations (\ref{burgers}) lead quite generically in the limit $\kappa
\rightarrow 0$ to the same picture: the dynamical evolution would be 
governed almost everywhere by Zel'dovich's approximation except at the
caustics, where the right--hand--side of Eq. (\ref{burgers}) would
dominate the evolution and a shock structure would be formed (known as
a ``pancake'' in the cosmological literature). Only the details of the
density and velocity profiles in the neighborhood of pancakes would
depend on $\gamma$.  It is an open question whether and in which cases
velocity dispersion and self--gravity could balance to form stable
soliton--like configurations.

\bigskip

\section {The Role of Noise: Emergence of a KPZ--like Equation} \label{KPZ}

In this section we consider the dynamical effect of the noise. We come
back to Eq. (\ref{euler2}) and introduce a velocity potential
$\psi ({\bf x}, t)$ by ${\bf u} \equiv - \nabla \psi$ (since the
parallelism assumption (\ref{parallelism}) implies that ${\bf u}$ is
irrotational). Then
\begin{multline}
\nabla \left[\frac{\partial \psi}{\partial t} - \frac{1}{2 a} 
(\nabla \psi)^2 + (H - F) \psi \right] = \\
= \frac{1}{a} \nabla \int^{\varrho}_{\varrho_{b}} dy \; 
\frac{p'(y)}{y} - {\bf s} \; .
\label{potential1}
\end{multline}

\noindent
We now make the assumption that the noise ${\bf s}$ is a potential
forcing, i.e., that there exists a (stochastic) potential $\eta$ such
that ${\bf s} \equiv - \nabla \eta$. This stochastic potential will be
chosen Gaussian--distributed with zero mean, $\langle \eta \rangle =
0$, and correlations
\begin{equation}
\langle \eta({\bf x}, t) \eta ({\bf x}', t') \rangle = 2  D ({\bf x}, 
{\bf x}', t, t') \; .
\end{equation}

\noindent
The assumption of potential noise can be motivated by the fact that in
the linear regime only the potential component of the noise
contributes to the growth of perturbations by gravitational
instability (see Sect. \ref{linear_regime} and Eq. (\ref{densityeq})
in Appendix B). If the noise had a non--potential component, it would
generate vorticity in the velocity field ${\bf u}$ and would
invalidate the parallelism assumption.

After inserting this stochastic force into Eq. (\ref{potential1}),
integrating, dropping an irrelevant additive function of time and
changing from $\varrho$ to $\delta$ in the integral, one obtains:
\begin{multline}
\frac{\partial \psi}{\partial t} - \frac{1}{2 a} (\nabla \psi)^2 + 
(H - F) \psi = \\
= \frac{1}{a} \int^{\delta}_{0} dy \; \frac{p'(\varrho_{b}(1+y))}{1+y} + 
\eta \; .
\label{potential2}
\end{multline}

\noindent
On the other hand, the constraint (\ref{constraint}) can be converted
into an equation relating the velocity potential $\psi$ and the
density contrast $\delta$, that is
\begin{equation}
\delta({\bf x}, t) = \frac{F(t)}{4 \pi G a(t) \varrho_{b}(t)} \nabla^2 
\psi ({\bf x}, t) \; .
\label{constraint2}
\end{equation}

\noindent
Inserting this equation into Eq. (\ref{potential2}) above then yields a
non--linear partial differential equation for the velocity
potential (after specifying the matter model $p(\varrho)$) given by:
\begin{multline}
\frac{\partial \psi}{\partial t} - \frac{1}{2 a} (\nabla \psi)^2 + 
(H - F) \psi = \\
=\frac{1}{a} \int^{\frac{F}{4 \pi G a \varrho_{b}} \nabla^2 \psi}_{0} 
dy \; \frac{p'(\varrho_{b}(1+y))}{1+y} + \eta \; .
\label{potential2bis}
\end{multline}

\noindent
There are two sources of non--linearities: the convective (also known
in the literature as advective) term $(\nabla \psi)^2$ and the
integral arising from the pressure force. To proceed further we need
an explicit expression for $p(\varrho)$ to evaluate the integral. For
a polytropic model $p=\kappa \varrho^{\gamma}$, one recovers Eq.
(\ref{closed_for_u}) but expressed in terms of the potential $\psi$
(plus the noise term). As discussed at the end of the previous
section, one may expect that the gross features of the LSS emerging
from these models are quite insensitive to the particular equation of
state $p=p(\varrho)$ in the limit of small pressure and that only the
fine details depend on it. We therefore simplify this equation further
by expanding the integral in a Taylor series and keeping only the
lowest order term in $\delta$. The resulting equation is {\it exact}
for the particular choice $p=\kappa \varrho^2$:
\begin{equation}
\frac{\partial \psi}{\partial t} - \frac{1}{2 a} (\nabla \psi)^2 + 
(H - F) \psi = \frac{F p'(\varrho_{b})}{4 \pi G a^2 \varrho_{b}} 
\nabla^2 \psi + \eta \; .
\label{potential3}
\end{equation}

\noindent
As remarked in the previous section, this choice leads to the
``adhesion approximation'', which is in fact postulated on a similar
reasoning: it is the simplest way of modelling ``sticking particles''
(Gurbatov et al. 1989). Hence, the ``adhesion approximation'' 
can be viewed as the first term in a Taylor series that approximates
the integral expression.

Eq. (\ref{potential3}) is the simplest equation for $\psi$ that
we can write, still containing the main ingredients that enter into the
physics of the self--gravitating gas we are describing, although as we
will see, it already entails a considerable degree of complexity.

\subsection{Discussion of Eq. (\ref{potential3})}

This stochastic partial differential equation is first
order in time and second order in position; it is also non--linear in
$\psi$ and the coefficients of the different terms are
functions of time. The physical meaning of each term is transparent: 

\noindent
(a) The term proportional to $\psi$ encompasses the competition
between damping of perturbations due to the cosmological expansion, ($H
\psi$), and enhancing of perturbations due to gravitational collapse,
($-F \psi$). This term introduces a time--dependent time scale
$|H-F|^{-1}$, which is the time scale for the damping (or enhancing)
of perturbations in regimes when the nonlinearity is negligible.

\noindent
(b) The term proportional to the Laplacian describes the
damping of perturbations due to velocity dispersion. This term
defines a time--dependent length scale, Jeans' length,
\begin{equation}
L_{J} := \left[ \frac{p'(\varrho_{b})}{4 \pi G a^2 \varrho_{b}} 
\right] ^{1/2} \; ,
\end{equation}

\noindent
which is discussed in Appendix B, after Eq. (\ref{Jeans_length}).

\noindent
(c) The non--linear term is the convective term in Euler's equation
(\ref{cosmohydrob}). The effect of this term is to broaden the peaks
in the field $\psi$. (An intuitive picture of how a term like $(\nabla
\psi)^2$ behaves may be found in Barab\'asi \& Stanley 1995,
 Fig. 6.2). 

\noindent
(d) The noise term incorporates the effects of fluctuations due to
different sources and leads to a roughening of the field $\psi$ in
space and in time as it evolves.

Eq. (\ref{potential3}) describes what is known in the condensed matter
literature as interface growth phenomena, but with two added
ingredients which are substantive to cosmology: the presence of
time--dependent coefficients due to cosmological expansion, and the
presence of a term proportional to $\psi$, which in the context of
condensed matter physics is interpreted as a finite, albeit
time--dependent, correlation length. This equation (Berera \& Fang
1994; Barbero et al. 1997) is then a generalization to cosmological
settings of the KPZ equation (Kardar et al. 1986, Barab\'asi \&
Stanley 1995) for surface growth.  In principle, one could use
techniques similar to those used there to study this equation, but it
turns out that complications arise as a consequence of the time
dependence of the coefficients. Because of this, the first thing one
thinks of is to perform changes of variables that will bring the
equation into an equation with constant coefficients from which one
can later on proceed with the analysis. With this in mind we rewrite
Eq. (\ref{potential3}) as
\begin{equation}
\frac{\partial\psi}{\partial t} = f_{1} (t) \nu \nabla^2 \psi + 
\frac{1}{2} f_{2} (t) \lambda (\nabla \psi)^2 + 
\frac{f_{3} (t)}{T} \psi + \eta \; ,
\label{potential4}
\end{equation}

\noindent
where we have defined the following dimensionless functions of time:
\begin{equation}
f_{1} (t) = \frac{1}{\nu} \frac{F(t) p'(\varrho_{b}(t))}
{4 \pi G a^2(t) \varrho_{b}(t)} \; ,
\label{f-functionsa}
\end{equation}
\begin{equation} 
f_{2} (t) = \frac{1}{\lambda a(t)} \; ,
\label{f-functionsb}
\end{equation}
\begin{equation}
f_{3} (t) = [F(t) - H(t)] T \; .  
\label{f-functionsc}
\end{equation}

\noindent
The dimensional parameters $\nu$, $\lambda$, $T$ are introduced as
bookkeeping quantities to carry the dimensions.  

Defining a new time coordinate $\tau$, a new velocity potential
$\Psi({\bf x}, \tau)$ and a new noise $\xi ({\bf x}, \tau)$ via
\begin{equation}
\tau(t) = \int^{t}_{t_{0}} dy f_1 (y) \; , \;\;\;\;\; 
\Psi= \frac{f_2}{f_1} \psi \; , \;\;\;\;\; 
\xi = \frac{f_{2}}{f_{1}^2} \eta \; ,
\end{equation}

\noindent
allows one to recast this equation into:
\begin{equation}
\frac{\partial \Psi}{\partial \tau} = \nu \nabla^2 \Psi + 
\frac{\lambda}{2} (\nabla \Psi)^2 - r(\tau) \Psi + \xi \; ,
\label{potential5}
\end{equation}

\noindent
where 
\begin{multline}
r(\tau) = \frac{1}{f_{1}(t(\tau))} \frac{d f_{1}(t(\tau))}{d \tau} -
\frac{1}{f_{2}(t(\tau))} \frac{d f_{2}(t(\tau))}{d \tau} - \\
- \frac{1}{T} \frac{f_{3}(t(\tau))}{f_{1}(t(\tau))} \; .
\label{linearcoefficient}
\end{multline}

The above equation is a ``massive'' KPZ equation, but with the
peculiarity that the coefficient of the term proportional to $\psi$
(the ``mass'' term) depends on time: one has a standard KPZ equation
if $r(\tau) = 0$, time--dependent damping of the surface growth if
$r(\tau)>0$, or explosive unstable behavior for $r(\tau)<0$.

Due to the noise term in Eq. (\ref{potential5}), the field $\Psi$ will
develop dynamical correlations in addition to those due to the initial
conditions. If the nonlinearity were neglected in Eq.
(\ref{potential5}), the effect of noise would simply be to superimpose
Gaussian fluctuations on the deterministic evolution.  But the
nonlinearity {\it couples different length scales}, so that the
evolution at any given scale also receives a contribution from
fluctuations on other scales. Eq. (\ref{potential5}) implicitly
involves a coarse-grained description and thus a smoothing length
scale, so that the nonlinear coupling promotes the phenomenological
coefficients $\nu$, $\lambda$, $r$ (and the noise itself) into
scale--dependent quantities.
  
The correlations induced by noise can be computed by means of the
Renormalization Group (Gell-Mann \& Low 1954, Binney et al. 1993,
Weinberg 1996), via the computation of the scale--dependence of the
coefficients. The fact that $r(\tau)$ depends on time again
complicates the application of the Renormalization Group. However, if
$r(\tau)$ happens to be independent of $\tau$ or if an adiabatic
approximation is justified (i.e., if $r(\tau)$ varies very slowly on
the time--scales associated with the dynamical evolution prescribed by
(\ref{potential5})), then one can to some degree of approximation
neglect the $\tau$--dependence of $r(\tau)$ when applying the
Renormalization Group. In Appendix A a detailed study of this question
is carried out.
  
Under the adiabatic assumption, application of the Renormalization
Group is straightforward albeit analytically cumbersome, and has been
carried out elsewhere (Barbero et al. 1997; Dom\'\i nguez et al.
1999). We here content ourselves by quoting the conclusions. The KPZ
equation exhibits self--affine correlations in the large-distance,
long-time regime, and the ``massive'' KPZ equation, Eq.
(\ref{potential5}), can also exhibit this behavior. In particular, the
equal-time two-point density correlation obeys in such case the
scaling $\langle \delta({\bf x}, t) \delta({\bf x}', t) \rangle \sim
|{\bf x}-{\bf x}'|^{2 \chi - 4}$, where $\chi$ is the ``roughness''
exponent, which can be computed by means of the Renormalization Group.
This result shows that noise may be relevant in that it can induce the
generation of self--affine correlations in a self--gravitating
collisionless gas, even if the initial conditions are not
self--affine.

\bigskip

\section {The Linear Theory Revisited} \label{linear_regime}

The strongest assumption that we have made has been the parallelism
hypothesis (\ref{parallelism}), motivated by the well--studied dust
case, where it is justified in the linear and weakly nonlinear
regimes. In this section we will study the linearized version of the
system of Eqs. (\ref{cosmohydroa}-\ref{cosmohydroc}) to find out how
well this assumption is justified in the presence of pressure and
noise.

The linearized Newtonian cosmological equations around the unperturbed
FL background (i.e., $\delta = 0$, ${\bf u = 0}$, ${\bf g = 0}$)
read:

\smallskip
\noindent
$\bullet$ Continuity equation:
\begin{equation}
\frac{\partial \delta}{\partial t} + \frac{1}{a} \nabla \cdot {\bf u} = 0 
\; ;
\label{linearcosmohydroa}
\end{equation}
$\bullet$ Euler's equation:
\begin{equation}
\frac{\partial {\bf u}}{\partial t} + H {\bf u} = {\bf g} - 
\frac{p'(\varrho_{b})}{a} \nabla \delta + {\bf s} \; ;
\label{linearcosmohydrob}
\end{equation}
$\bullet$ Newtonian field equations:
\begin{equation} 
\nabla \cdot {\bf g} = - 4 \pi G a \varrho_{b} \delta
\; , \;\;\;\;\; \nabla \times {\bf g} = 0 \; .
\label{linearcosmohydroc}
\end{equation}

Since the noise ${\bf s}({\bf x}, t)$ has vanishing mean and the
equations are linear, the general solution can be split into the sum
of the averaged fields $\langle \delta \rangle$, $\langle {\bf u}
\rangle$ and $\langle {\bf g} \rangle$, obeying the deterministic
version of the Eqs. (\ref{linearcosmohydroa}-\ref{linearcosmohydroc})
(i.e., setting ${\bf s}={\bf 0}$), plus a fluctuating part which is a
linear functional of the noise ${\bf s}$. The dynamical evolution thus
follows a deterministic trajectory with Gaussian fluctuations
superimposed on it, but these corrections will be small if the noise
intensity is small. The detailed study of this system of equations is
performed in Appendix B. The main conclusions for our purposes are the
following:

(i) The average vorticity of the peculiar--velocity is damped by the
background expansion, even if $p \neq 0$, and thus $\langle {\bf u}
\rangle$ becomes asymptotically a potential field as $t \rightarrow +
\infty$.

(ii) The parallelism assumption (\ref{parallelism}) gets modified due
to pressure. When the lowest order correction is included, we find
that \footnote{The notation and the assumptions leading to this
expression are explained in Appendix B.}
\begin{equation}
\langle {\bf g} \rangle \approx F_{0}(t) \langle {\bf u} \rangle 
- \frac{p'(\varrho_{i}) F_{1}(t)}{4 \pi G a^2_{i} \varrho_{i}} 
\nabla^2 \langle {\bf u} \rangle \; .
\label{newparallelism}
\end{equation} 

\noindent
This equation shows that corrections to parallelism in the average
motion are negligible in the limit of small pressure far from caustics
(i.e., where $\langle {\bf u} \rangle$ is a smooth field and $\nabla^2
\langle {\bf u} \rangle$ does not diverge). Pressure becomes important
in pancakes, where parallelism no longer holds and also the
assumptions of small and isotropic pressure break down.

\bigskip

\section{Conclusions} \label{conclusions}

In this paper we have put forward the set of Eqs.
(\ref{cosmohydroa}-\ref{cosmohydroc}) as a new, phenomenological
approach to the problem of LSS formation. The difference with regard
to similar approaches lies in the interpretation of these equations as
describing the dynamical evolution of the {\em coarse--grained} fields
$\delta$, ${\bf u}$ and ${\bf g}$, which is the origin of the
pressure--like force and of the fluctuations (noise).

We have shown that under the parallelism assumption
(\ref{parallelism}), the pressure force gives rise to a viscous--like
force of the same character as that of the ``adhesion model'', which
is a successful model on large scales. As is also known in the context
of the ``adhesion model'', the limit of vanishing pressure is singular:
the models with no pressure ($p = 0$ exactly) are {\em qualitatively}
different from the models with $p \neq 0$. Therefore, even if pressure
is very small (seemingly negligible), one should not set $p=0$.

We have also considered the effects of fluctuations. Under the
parallelism assumption, the problem can be cast into a ``massive'' KPZ
model of surface growth phenomena, which can exhibit self--affine
correlations because the noise can be relevant, even if it is
vanishingly small. Hence, the limit of vanishing noise is also
singular and the same warning as with pressure applies.

Finally, we have explored the plausibility of the parallelism
assumption (\ref{parallelism}) by studying the linearized equations
and reached the conclusion that it is justified far from caustics and
close to the limit of vanishing pressure and noise. In fact, given the
number of assumptions we have made, there is plenty of room to
generalize the results presented here: relaxing the parallelism
assumption, generalizing the equation of state $p=p(\varrho)$,
modelling multi--streaming by an anisotropic stress--tensor $\Pi_{ij}$
in Eq. (\ref{cosmohydrob}) rather than by a pressure force (see, e.g.,
Maartens et al. 1999), as well as performing a full Renormalization
Group analysis with time--dependent coefficients.

Furthermore, we have given the formal hydrodynamical basis on which to
anchor future studies leading to an understanding of various scaling
relations found in the context of LSS.

\begin{acknowledgements}
  
  We would like to thank Herbert Spohn for valuable discussions.
  Thomas Buchert acknowledges support by the ``Sonderforschungsbereich
  SFB 375 f\"ur Astro--Teilchenphysik der Deutschen
  Forschungsgemeinschaft'' and thanks LAEFF for hospitality and
  generous support during working visits in Madrid. Alvaro Dom\'\i
  nguez thanks ``Sonderforschungsbereich SFB 375'' for support during
  working visits at the Ludwig--Maximilians--Universit\"at.

\end{acknowledgements}

\appendix

\section*{Appendix A: Adiabatic approximation}

In this appendix we explore under what conditions the coefficient
$r(\tau)$ in Eq. (\ref{potential5}) can be considered almost
time--independent. Our starting point is Eq.
(\ref{linearcoefficient}):
\begin{multline}
r(\tau) = \frac{1}{f_{1}(t(\tau))} \frac{d f_{1}(t(\tau))}{d \tau} -
\frac{1}{f_{2}(t(\tau))} \frac{d f_{2}(t(\tau))}{d \tau} - \\
- \frac{f_{3}(t(\tau))}{T f_{1}(t(\tau))} \; .
\label{rcoeff1}
\end{multline}

\noindent
The function $F$ in the definitions
(\ref{f-functionsa}-\ref{f-functionsc}) obeys Eq.
(\ref{Fexpansiona}) of Appendix B. Therefore:
\begin{equation}
\frac{1}{f_{1}} \frac{d f_{1}}{d t} = \frac{4 \pi G \varrho_{b}}{F}
 - F - 3 H \varrho_{b} \frac{p''(\varrho_{b})}{p'(\varrho_{b})} 
\; ,
\end{equation}
\begin{equation}
\frac{1}{f_{2}} \frac{d f_{2}}{d t} = - H \; .
\end{equation}

\noindent
Taking these results into Eq. (\ref{rcoeff1}) yields
\begin{equation}
r = \frac{4 \pi G \varrho_{b} a^2}{F p'(\varrho_{b})} \left[ 
\frac{4 \pi G \varrho_{b}}{F} + 2 (H - F) - 3 H \varrho_{b} 
\frac{p''(\varrho_{b})}{p'(\varrho_{b})} \right] \; .
\label{rcoeff2}
\end{equation}

\noindent
On the other hand, one can identify two time scales in Eq.
(\ref{potential5}): a time scale due to expansion
\begin{equation}
T_{exp} := a \left( \frac{d a}{d \tau} \right)^{-1} = 
\frac{f_{1}}{H} \; ,
\end{equation}

\noindent
and an intrinsic time scale $T_{int} := |r|^{-1}$. If the {\em
relative} variation of the coefficient $r(\tau)$ on these time
scales is small, then an adiabatic approximation is justified and $r$
can be considered to be a slowly varying function of time. To
quantify this statement we define {\it adiabatic indices} as
\begin{equation}
E_{\exp} := T_{exp} \left| \frac{1}{r} \frac{d r}{d \tau} \right| =
\left| \frac{1}{H r} \frac{d r}{d t} \right| \; ,
\end{equation}
\begin{equation}
E_{int} := T_{int} \left| \frac{1}{r} \frac{d r}{d \tau}
\right| = \left| \frac{1}{f_{1} r^2} \frac{d r}{d t} \right|
\; .
\label{adiabatic_indices}
\end{equation}

\noindent
The adiabatic approximation then requires $E_{exp}$, $E_{int} \ll
1$. We now discuss the various background cosmologies.

\subsection*{A.1. Einstein--de Sitter background ($K=0$, $\Lambda=0$)}

We first study the simplest case, for which
\begin{equation}
a(t) = \left( \frac{t}{t_{0}} \right)^{2/3} , \; 
H(t) = \frac{2}{3 t} , \; \varrho_{b} (t) = 
\frac{1}{6 \pi G t^2} .
\end{equation}

\noindent
From Eq. (\ref{flatF0}) of Appendix B we find that
$F=t^{-1}$. Taking these expressions into Eq. (\ref{rcoeff2})
then yields
\begin{equation}
r(t) = - \frac{2}{9 \pi G t_{0}^{4/3}} \frac{p''(\varrho_{b}(t))}
{p'(\varrho_{b}(t))^2} \, t^{-8/3} \; .
\end{equation}

\noindent
We are interested in polytropic models $p= \kappa
\varrho^{\gamma}$, for which the previous expression becomes
\begin{equation}
r(t) = \frac{4 (6 \pi G)^{\gamma-1} (1 - \gamma)}
{3 \kappa \gamma t_{0}^{4/3}} \, t^{\frac{2}{3} (3 \gamma - 4)} \; .
\end{equation}

\noindent
We find some interesting conclusions: for the isothermal model,
$\gamma=1$, we get $r=0$ {\em exactly} and Eq. (\ref{potential5})
reduces to a {\it bona fide} KPZ equation. For the cosmogenetic model
$\gamma=4/3$, we get $r(t) = r_{0}<0$ independent of time. For other
matter models we have from Eq. (\ref{adiabatic_indices}):
\begin{equation}
E_{exp} = |3 \gamma - 4| \; , \;\;\;\;\; E_{int} = 
\frac{|3 \gamma - 4|}{3 |\gamma - 1|} \; .
\end{equation}

\noindent
Then the adiabatic approximation will be justified when $\gamma$ is
close to the cosmogenetic value of $4/3$, in which case $r<0$.

\subsection*{A.2. Background cosmologies with ($K \neq 0$, $\Lambda = 0$)}

In this case we can also obtain explicit expressions for the
coefficient $r$ and the indices $E_{exp}$, $E_{int}$ analytically, but
the algebra is rather involved and the final expressions are best
studied numerically. We therefore skip the algebra and go right to the
final results, particularized to the isothermal model $p =
\kappa \varrho$, because this leads to the simplest expression in
Eq. (\ref{rcoeff2}).

\begin{figure}
  \resizebox{\hsize}{!}{\includegraphics[width=9cm]{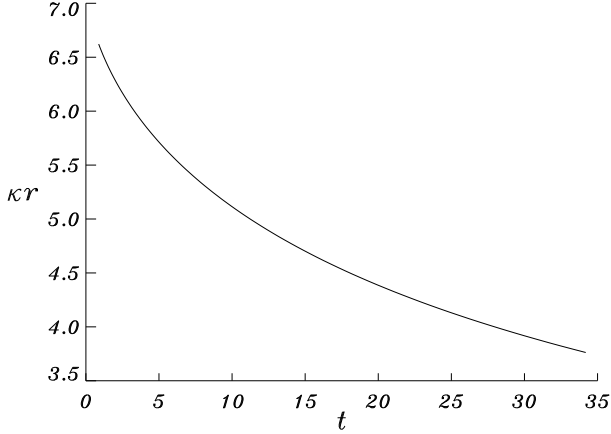}}
  \caption{$\kappa r$ versus time in an undercritical FL background 
    ($\Omega_{0}<1$). $\kappa r$ is measured in units of $ 10^{-3} 
    (1 - \Omega_0) h^{2} \, Gyr^{-2}$, and time is measured in units of 
    $\Omega_{0} (1 - \Omega_{0})^{-3/2} h^{-1} \, Gyr$.}
  \label{openr_t}
\end{figure}

In an undercritical universe ($K<0$) the coefficient $r$ is a
positive, monotonically decreasing function of time. In
Fig.~\ref{openr_t} we plot $\kappa r$ versus $t$ \footnote{We have
  introduced Hubble's constant at the present epoch, $H_{0} = 0.1 h \,
  Gyr^{-1}$, and the density parameter, $\Omega_{0} = 8 \pi G
  \varrho_{0} / 3 H_{0}^2$.}. The adiabatic indices are monotonically
increasing functions of time but they are bounded:
\begin{equation}
0 \leq E_{exp} \leq 1 \; , \;\;\;\;\; \frac{32}{75} \leq E_{int} 
\leq \frac{1}{2} \; .
\end{equation}

\noindent
In the range $0.2 \lesssim \Omega_{0} \leq 1$ and at epochs $t
\lesssim 10 h^{-1} \, Gyr$ one gets $0 \leq E_{exp} \lesssim
0.35$. Although the adiabatic indices are not much smaller than one,
one can still work under the adiabatic assumption to a first
approximation.

\medskip

\begin{figure}
 \resizebox{\hsize}{!}{\includegraphics[width=9cm]{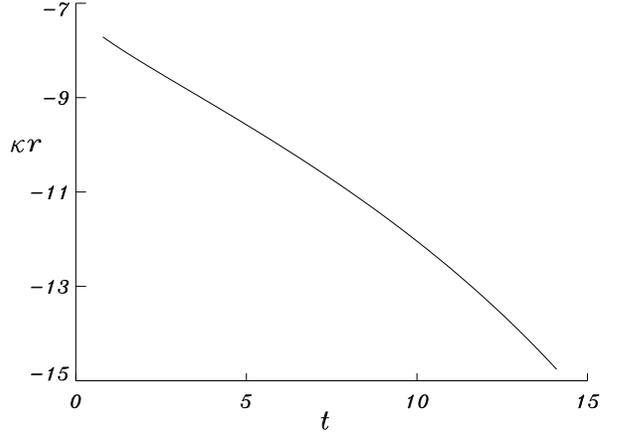}}
 \caption{$\kappa r$ versus time in an overcritical FL background 
   ($\Omega_{0}>1$) up to the epoch of maximum expansion. $\kappa r$ is
   measured in units of $ 10^{-3} (\Omega_0 - 1) h^{2} \, Gyr^{-2}$, and 
   time is measured in units of 
   $\Omega_{0} (\Omega_{0} - 1)^{-3/2} h^{-1} \, Gyr$.}
 \label{closedr_t}
\end{figure}

In an overcritical universe ($K>0$) the coefficient $r$ is a negative,
decreasing function of time. In Fig.~\ref{closedr_t} we plot $\kappa
r$ versus $t$. The index $E_{int}$ decreases in time and is bounded:
\begin{equation}
\frac{2 \pi^2 (9 \pi^2 - 64)}{(3 \pi^2 -64)^2} \leq 
E_{int} \leq \frac{32}{75} \; .
\end{equation}

\noindent
The index $E_{exp}$ increases with time and is unbounded: it diverges
at the epoch of maximum expansion because then the expansion rate $H$
vanishes. This is, however, physically uninteresting. At epochs $t
\lesssim 10 h^{-1} \, Gyr$ and with the conservative bound $\Omega_{0}
\lesssim 2$, one finds $E_{int} \lesssim 0.5$ and the same remarks
apply as in the undercritical universe.

\medskip

Therefore we conclude that within the physically interesting range
of values of the parameters, an adiabatic approximation for the
coefficient $r$ may be justified.

\bigskip

\section*{Appendix B: Linear regime} \label{ap_lin}

In this appendix we study the linear regime when pressure is taken
into account. The linearized system of Eqs.
(\ref{linearcosmohydroa}-\ref{linearcosmohydroc}) can be more easily
studied in Fourier space. We therefore define the Fourier
transformation (assuming a box of volume $\cal V$):
\begin{multline}
\delta ({\bf x}, t) = \frac{1}{{\cal V}} \sum_{{\bf k}} \tilde{\delta}
({\bf k}, t) e^{-i {\bf k} \cdot {\bf x}} \;\; \Rightarrow \\
\Rightarrow \;\; \tilde{\delta} ({\bf k}, t) = \int_{\cal V} d^{3} 
{\bf x} \; \delta ({\bf x}, t) e^{i{\bf k} \cdot {\bf x}} \; ,
\end{multline}

\noindent
and similar definitions for $\tilde{\bf u} ({\bf k}, t)$,
$\tilde{\bf g} ({\bf k}, t)$ and $\tilde{\bf s} ({\bf k}, t)$. In
Fourier space the linearized Eqs.
(\ref{linearcosmohydroa}-\ref{linearcosmohydroc}) read: 

\noindent
\begin{equation}
\frac{\partial \tilde{\delta}}{\partial t} - \frac{i}{a} {\bf k} \cdot
\tilde{{\bf u}} = 0 \; ;
\label{fouriera}
\end{equation}
\begin{equation}
\frac{\partial \tilde{\bf u}}{\partial t} + H \tilde{\bf u} = \tilde{\bf g}
+ i \frac{p'(\varrho_{b})}{a} \, \tilde{\delta} \, {\bf k} + \tilde{\bf s} \; ;
\label{fourierb}
\end{equation}
\begin{equation}
i {\bf k} \cdot \tilde{\bf g} = 4 \pi G a \varrho_{b} \tilde{\delta}
\; , \;\;\;\;\; {\bf k} \times \tilde{\bf g} = {\bf 0} \; .  
\label{fourierc}
\end{equation}

We decompose the velocity field in transversal $\tilde{{\bf u}}_{\perp}$ 
(satisfying ${\bf k} \cdot \tilde{{\bf u}}_{\perp}=0$) and
longitudinal components $\tilde{{\bf u}}_{\parallel}$ (satisfying ${\bf k}
\times \tilde{{\bf u}}_{\parallel}={\bf 0}$), and the same with the  
fields ${\tilde {\bf g}}$ and $\tilde{\bf s}$. From Eq. (\ref{fourierc}) 
one finds that ${\tilde {\bf g}}_{\perp}={\bf 0}$ and then from Euler's 
equation (\ref{fourierb}) that:
\begin{multline}
\frac{\partial \tilde{{\bf u}}_{\perp}}{\partial t} + H 
\tilde{{\bf u}}_{\perp} = \tilde{\bf s}_{\perp} \;\; \Rightarrow \\
\Rightarrow \;\; \tilde{{\bf u}}_{\perp} ({\bf k}, t) = \frac{1}{a(t)} [ 
{\tilde {\bf u}}_{\perp} ({\bf k}, t_{0}) + \int_{t_0}^t d \tau \; 
a(\tau) \tilde{\bf s} ({\bf k}, \tau) ] \; .
\end{multline}

\noindent
Hence, the average transversal component of the peculiar--velocity is
damped by the background expansion even in the presence of pressure
(this is not surprising, since pressure does not generate vorticity
for barotropic fluids, i.e., $p=p(\varrho)$), and only its
fluctuations grow due to the noise. The longitudinal component is
obtained from the continuity equation (\ref{fouriera}), while the
gravitational acceleration is found from Poisson's equation
(\ref{fourierc}):
\begin{equation}
\tilde{{\bf u}}_{\parallel} = - \frac{i a}{k^2} \frac{\partial
\tilde{\delta}}{\partial t} {\bf k} \; , \;\;\;\;\; 
\tilde{{\bf g}} = - \frac{4 \pi G i a \varrho_{b}}{k^2} \tilde{\delta} 
{\bf k} \; .
\label{uw(delta)}
\end{equation}

\noindent
The elimination of the velocity and gravitational acceleration fields
in favor of the density contrast field by means of these expressions
yields a closed equation for each mode $\tilde{\delta}({\bf k}, t)$ of
the density contrast:
\begin{equation}
\frac{\partial^{2} \tilde{\delta}({\bf k}, t)}{\partial t^{2}} + 2 H
\frac{\partial \tilde{\delta}({\bf k}, t)}{\partial t} - U_{k}
\tilde{\delta}({\bf k}, t) = \frac{1}{a(t)} i {\bf k} \cdot 
\tilde{\bf s}_{\parallel} ({\bf k}, t) \; ,
\label{densityeq}
\end{equation}
with
\begin{equation}
U_{k} := 4 \pi G \varrho_{b}- \frac{k^{2}}{a^{2}} p'(\varrho_{b}) \; ,
\end{equation}

\noindent
It is clear from this equation that the noise will induce correlations
on the density field on top of those already present in the initial
condition. In the linear regime, this superposition is linear and weak
noise will only induce small Gaussian fluctuations of $\tilde{\delta}$
about the deterministic solution. Hence, we can restrict our study to
the evolution of the average field, $\langle \tilde{\delta} \rangle$,
which follows the deterministic evolution (i.e., Eq. (\ref{densityeq})
dropping the noise). This evolution is, however, still different from
the dust case because of the pressure term. In the rest of the
appendix we drop the noise source from this equation and understand
that $\tilde{\delta}$ really means $\langle \tilde{\delta} \rangle$,
and the same with the other fields.

We are particularly interested in the justification of the parallelism
condition, Eq. (\ref{parallelism}). From Eqs. (\ref{uw(delta)}) one
can easily find a relationship between $\tilde{\bf u}_{\parallel}$ and
$\tilde{\bf g}$:
\begin{equation}
\tilde{\bf g}({\bf k}, t) = {\tilde F}({\bf k}, t) \; 
\tilde{\bf u}_{\parallel} \;, 
\label{fourierparallelism}
\end{equation}
where 
\begin{equation} 
{\tilde F} ({\bf k}, t) = 4 \pi G \varrho_{b}(t) \; \tilde{\delta} 
({\bf k}, t) \left(\frac{\partial \tilde{\delta} ({\bf k}, t)}{\partial t} 
\right)^{-1} \; .
\label{Fdef} 
\end{equation}

\noindent
The equation satisfied by ${\tilde F}$ can be easily found from 
Eq. (\ref{densityeq}):
\begin{equation}
\frac{\partial \tilde {F}}{\partial t} = 4 \pi G \varrho_{b} - 
H {\tilde F} - \frac{U_{k}}{4 \pi G \varrho_{b}} {\tilde F}^2 \; , 
\label{Feq}
\end{equation}

\noindent
(i.e., a Ricatti equation). There is parallelism in position space
between ${\bf u}$ and ${\bf g}$ if $\tilde F$ does not depend on ${\bf
k}$; a necessary condition is that $U_{k}$ be also $k$--independent
(i.e., there is no pressure). Therefore, not surprisingly, pressure
destroys parallelism, but the deviations decrease as one goes to
larger scales.

The exact solution to Eq. (\ref{Feq}) is difficult to obtain. For
polytropic models in an Einstein--de Sitter background ($K=0$,
$\Lambda=0$) one can solve Eq. (\ref{densityeq}) in terms of Meijer's
G--functions (Haubold et al. 1991) and then compute $\tilde F$ from
its definition, Eq. (\ref{Fdef}). This procedure is, however,
algebraically cumbersome and not very illuminating from the physical
point of view (for some particular values of the polytropic index
$\gamma$ however the solution can be written in terms of elementary
functions).

For the purposes of the work presented here, we adopt a different
approach. We first define Jeans' length $L_{J}$ by the condition
$U_{L_{J}^{-1}}=0$, yielding
\begin{equation}
L_{J} := \left[ \frac{p'(\varrho_{b})}{4 \pi G a^2 \varrho_{b}}
\right]^{1/2} \; .
\label{Jeans_length}
\end{equation}

\noindent
From Eq. (\ref{densityeq}) one can easily grasp the physical meaning of
this quantity: density perturbations with $k>L_{J}^{-1}$ are damped in
the linear regime by both pressure and expansion, while those with
$k<L_{J}^{-1}$ are damped only by expansion (self--gravity dominates
over pressure).

In the limit of small pressure, $L_{J} \rightarrow 0$, and the
pressure can be considered a perturbation on scales $k \ll
L_{J}^{-1}$. Defining the small parameter $\varepsilon := L_{i}^2 k^2$
(where $L_{i}$ is Jeans' length evaluated at some initial time
$t_{i}$), we can write
\begin{equation}
\frac{U_{k}}{4 \pi G \varrho_{b}} = 1 - \varepsilon S(t) \; , \;\;\; 
S(t) := \frac{L_{J}^2}{L_{i}^2} = \frac{a}{a_{i}} 
\frac{p'(\varrho_{b})}{p'(\varrho_{i})} \; ,
\end{equation}

\noindent
and then try a perturbative expansion in powers of $\varepsilon$. In
particular, for polytropic models one has $S(t) = [a(t)/a_{i}]^{4 - 3
\gamma}$, that is, the perturbative expansion is better at later times if 
$\gamma>4/3$. It should also be noticed that the correction due to
pressure at a fixed time becomes smaller as the polytropic index
$\gamma$ grows.

We therefore write ${\tilde F}({\bf k}, t) = \sum_{n=0}^{\infty}
\varepsilon ^n F_{n}({\bf k}, t)$ with the initial conditions
$F_{0}({\bf k}, t_{i}) = {\tilde F} ({\bf k}, t_{i})$ and $F_{n}({\bf
k}, t_{i}) = 0$, $n>0$. From Eq. (\ref{Feq}) one finds
\begin{equation}
\frac{\partial F_{0}}{\partial t} = 4 \pi G \varrho_{b} - H F_{0} 
- F_{0}^2 \; ,
\label{Fexpansiona}
\end{equation}
\begin{equation}
\frac{\partial F_{1}}{\partial t}= - (H + 2 F_{0}) F_{1} + S F_{0}^2 \; , 
\label{Fexpansionb}
\end{equation}
\begin{displaymath}
\vdots 
\end{displaymath}

\noindent
$F_{0}$ obeys the equation of the well--studied pressureless case
(Peebles 1980). As $t \rightarrow \infty$ it becomes independent of
$k$ (i.e., it forgets initial conditions) and there is parallelism in
position space. The asymptotic form of $F_{0}$ is to be identified
with $F(t)$ in the parallelism assumption (Eq. [\ref{parallelism}] in
the main text), and turns out to be given as: $F_{0} = 4 \pi G
\varrho_{b} b / {\dot b}$, where $b(t)$ is the {\em growing} solution
of Eq. (\ref{densityeq}) when particularized to the dust case,
$p=0$. $F_{1}$ can be computed once $F_{0}$ is known:
\begin{equation}
F_{1}({\bf k}, t) = \int_{t_{i}}^{t} d \tau \; S(\tau) F_{0}^2({\bf
k}, \tau) e^{- \int_{\tau}^{t} d \theta \; [ H(\theta) + 2 F_{0}({\bf
k}, \theta) ] } \; .
\label{F1}
\end{equation}

\noindent
Notice that the whole dependence of $F_{1}$ on ${\bf k}$ stems from
$F_{0}$ (and hence from initial conditions, which is true for every
$F_{n}$ too). The study of this expression can be carried out
analytically for polytropic models, $p = \kappa \varrho^{\gamma}$, in
an Einstein--de Sitter background. In this case the function $F_{0}$
is
\begin{equation}
F_{0}({\bf k}, t) = \frac{t^{5/3} + 2 A({\bf k})}{t^{8/3} - 3 
A({\bf k}) t} \; ,
\label{flatF0}
\end{equation}

\noindent
with $A({\bf k})$ given by the initial conditions. For polytropic
models the function $S(t)$ comes down to the simple expression $S(t) =
(t/t_{i})^{8/3 - 2 \gamma}$. Taking these results into Eq.
(\ref{F1}) then yields
\begin{equation}
F_{1}({\bf k}, t) = \frac{t_{i}^{-\frac{8}{3} + 2 \gamma}
t^{2/3}}{[t^{5/3} - 3 A({\bf k})]^2} \bigg[\frac{3}{13 - 6 \gamma}
(t^{\frac{13}{3} - 2 \gamma} - t_{i}^{\frac{13}{3} - 2 \gamma}) + 
\nonumber
\end{equation}
\begin{equation}  
 + \frac{12 A({\bf k})}{8 - 6 \gamma} (t^{\frac{8}{3} - 2 \gamma} 
- t_{i}^{\frac{8}{3} - 2 \gamma}) + \frac{A({\bf k})^2}{1 - 2 \gamma} 
(t^{1 - 2 \gamma} - t_{i}^{1 - 2 \gamma}) \bigg] \; ,
\end{equation}

\noindent
with the understanding that terms like $(t^a - t_{i}^a)/a$ should be
substituted by $\log (t/t_{i})$ when $a=0$.

As $t \rightarrow + \infty$, one has the asymptotic behaviors
$F_{0}({\bf k}, t) \rightarrow t^{-1}$ and
\begin{equation}
F_{1}({\bf k}, t) \rightarrow \frac{3}{(13 - 6 \gamma) t_{i}} 
\left(\frac{t}{t_{i}}\right)^{\frac{5}{3} - 2 \gamma} \; , 
\;\;\;\;\; (\gamma \neq \frac{13}{6}) \; ,
\end{equation}
\begin{equation}
F_{1}({\bf k}, t) \rightarrow \frac{1}{t_{i}} \left(\frac{t}{t_{i}}
\right)^{-\frac{8}{3}} \log \left(\frac{t}{t_{i}}\right) \; ,
\;\;\;\;\; (\gamma = \frac{13}{6}) \; .
\end{equation}

\noindent
It must be noted that $F_{0}$ and $F_{1}$ forget initial conditions
and thus become asymptotically $k$--independent, so that Eq.
(\ref{fourierparallelism}) yields
\begin{equation}
\tilde{\bf g} ({\bf k}, t) = [ F_{0}(t) + k^2 L_{i}^2 
F_{1}(t) + o(k^2 L_{i}^2)^2 ] \; \tilde{\bf u}_{\parallel} ({\bf k}, t) 
\; .
\end{equation}

\noindent
Assuming that the velocity field ${\bf u}$ is smooth (and in
particular not vertical or multi--valued), such that $\tilde{\bf u}$
decays fast enough when $k \rightarrow \infty$, we can Fourier
transform this expression back to position space and thus obtain the
following relationship between gravitational acceleration and the
potential component of the velocity,
\begin{equation}
{\bf g} \approx F_{0}(t) \; {\bf u}_{\parallel} - L_{i}^2 F_{1}(t) 
\nabla^2 {\bf u}_{\parallel} \; ,
\end{equation}

\noindent
which is the result quoted in Sect. \ref{linear_regime}.




\end{document}